%% file: main.tex
\documentclass[sigconf]{acmart}

\usepackage{float}
\usepackage{xcolor}
\usepackage{multirow}
\usepackage{paralist}
\usepackage{booktabs}
\usepackage{xspace}
\usepackage{subcaption}
\usepackage{balance} 

\definecolor{black}{rgb}{0,0,0}
\definecolor{grey}{rgb}{0.8,0.8,0.8}
\definecolor{red}{rgb}{1,0,0}
\definecolor{green}{rgb}{0,1,0}
\definecolor{darkgreen}{rgb}{0,0.5,0}
\definecolor{darkpurple}{rgb}{0.5,0,0.5}
\definecolor{darkdarkpurple}{rgb}{0.3,0,0.3}
\definecolor{blue}{rgb}{0,0,1}
\definecolor{shadegreen}{rgb}{0.95,1,0.95}
\definecolor{shadeblue}{rgb}{0.95,0.95,1}
\definecolor{shadered}{rgb}{1,0.85,0.85}
\definecolor{shadegrey}{rgb}{0.85,0.85,0.85}
\definecolor{oddRowGrey}{rgb}{0.80,0.80,0.80}
\definecolor{evenRowGrey}{rgb}{0.85,0.85,0.85}

\definecolor{ForestGreen}{rgb}{0.0, 0.66, 0.47}
\definecolor{RubineRed}{rgb}{1.0, 0.0, 0.31}

\newtheorem{Problem}{Problem}

\newcommand{\benchmark}{\ensuremath{\mathsf{Machamp}}\xspace}
\newcommand{\svm}{\textsf{SVM}\xspace}
\newcommand{\rf}{\textsf{Random Forest}\xspace}
\newcommand{\ditto}{\textsf{Ditto}\xspace}
\newcommand{\transf}{\textsf{Transformer}\xspace}
\newcommand{\dm}{\textsf{Deep Matcher}\xspace}
\newcommand{\der}{\textsf{DeepER}\xspace}
\newcommand{\sbert}{\textsf{SentenceBERT}\xspace}

\newcommand{\rrh}{\textsc{Rel-HETER}\xspace}
\newcommand{\sso}{\textsc{Semi-HOMO}\xspace}
\newcommand{\ssh}{\textsc{Semi-HETER}\xspace}
\newcommand{\sr}{\textsc{Semi-Rel}\xspace}
\newcommand{\stw}{\textsc{Semi-Text-w}\xspace}
\newcommand{\stc}{\textsc{Semi-Text-c}\xspace}
\newcommand{\rt}{\textsc{Rel-Text}\xspace}

\usepackage[shortlabels]{enumitem}
\setlist[itemize]{leftmargin=*}
\setlist[enumerate]{leftmargin=*}

\usepackage[linesnumbered,ruled,vlined]{algorithm2e}

\usepackage{dsfont}

\usepackage{multirow}
\usepackage[font=small,skip=3pt]{caption}

\AtBeginDocument{%
  \providecommand\BibTeX{{%
    \normalfont B\kern-0.5em{\scshape i\kern-0.25em b}\kern-0.8em\TeX}}}

\setcopyright{acmcopyright}
\copyrightyear{2021}
\acmYear{2021}
\acmConference[CIKM '21]{CIKM '21: 30th ACM International Conference on Information and Knowledge Management}{November 1-5, 2021}{Online}
\acmBooktitle{CIKM '21: 30th ACM International Conference on Information and Knowledge Management}

\begin{document}

\title{\benchmark: A Generalized Entity Matching Benchmark}

%

\author{Jin Wang}
\email{jin@megagon.ai}
\affiliation{%
  \institution{Megagon Labs}
  \country{United States}
}

\author{Yuliang Li}
\email{yuliang@megagon.ai}
\affiliation{%
  \institution{Megagon Labs}
  \country{United States}
}

\author{Wataru Hirota}
\email{wataru@megagon.ai}
\affiliation{%
  \institution{Megagon Labs}
  \country{United States}
}

\renewcommand{\shortauthors}{Wang et al.}

\begin{abstract}
	Entity Matching (EM) refers to the problem of determining whether two different data representations 
	refer to the same real-world entity.
	It has been a long-standing interest of the data management community and many efforts have been paid in creating benchmark tasks as well as in developing advanced matching techniques.
	However, existing benchmark tasks for EM are limited to the case where 
	the two data collections of entities are structured tables with the same schema.
	Meanwhile, the data collections for matching could be structured, semi-structured, or unstructured in real-world scenarios of data science.
	In this paper, we come up with a new research problem -- Generalized Entity Matching to satisfy this requirement and create a benchmark \benchmark for it.
	\benchmark consists of seven tasks having diverse characteristics and thus provides good coverage of use cases in real applications.
	We summarize existing EM benchmark tasks for structured tables and conduct a series of processing and cleaning efforts to transform them into matching tasks between tables with different structures.
	Based on that, we further conduct comprehensive profiling of the proposed benchmark tasks and evaluate popular entity matching approaches on them.
	With the help of \benchmark, it is the first time that researchers can
	evaluate EM techniques between data collections with different structures. 
\end{abstract}


\maketitle

\input{sec1-intro}
\input{sec2-tasks}
\input{sec3-datasets}
\input{sec4-experiment}
\input{sec5-related}

\input{sec6-conclude}

\balance
\bibliographystyle{ACM-Reference-Format}
\bibliography{main}
\appendix
\input{sec-schema}

\end{document}

%% file: sec1-intro.tex
\section{Introduction} \label{sec:intro}

Given two collections of entity records, the goal of Entity Matching (EM) is to identify pairs of records
that refer to the same real-world entity. Entity Matching is also known as Entity Resolution, Record Linkage, or Entity De-duplication.
As a fundamental problem in data integration, EM has a wide range of applications from data cleaning, knowledge base construction to entity clustering and search~\cite{fellegi69,DBLP:journals/tkde/Christen12,DBLP:books/daglib/0029346,DBLP:journals/pvldb/KondaDCDABLPZNP16,DBLP:journals/jdiq/LiLSWHT21}.
More recently, machine learning techniques, particularly deep learning with pre-trained language 
models~\cite{DBLP:journals/pvldb/KondaDCDABLPZNP16,DBLP:journals/pvldb/EbraheemTJOT18,DBLP:conf/sigmod/MudgalLRDPKDAR18,DBLP:journals/pvldb/EbraheemTJOT18,DBLP:conf/edbt/BrunnerS20,DBLP:journals/pvldb/0001LSDT20}, achieved the state-of-the-art (SOTA) matching quality among EM tasks.

Although significant progress has been made, 
existing EM solutions impose several assumptions that limit them to be applied to more practical scenarios.
First, they typically assume that the entity records are stored in a 
structured format such as relational tables.
In real scenarios, however, entities can be represented in a diverse set of data formats.
For example, job postings in recruit platforms are stored in semi-structured JSON and product reviews in e-commerce websites are stored in plain-text documents.
A direct application of existing methods would require flattening the nested attributes which can lead to
a potential loss of important structural information such as the list of job categories attached to the job posting.
Second, EM solutions assume the two entity collections to have identical or aligned schema so that
they can compute attribute-wise similarity scores as matching features \cite{DBLP:journals/pvldb/KondaDCDABLPZNP16,DBLP:journals/pvldb/EbraheemTJOT18,DBLP:conf/sigmod/MudgalLRDPKDAR18}.
To Impose this assumption, we need a potentially expensive schema matching in the pre-processing
step~\cite{DBLP:journals/pvldb/BernsteinMR11}, which is even not applicable when matching data of \emph{heterogeneous format}, 
e.g. matching product meta-information stored in tabular format with their textual descriptions.
Finally, traditional EM tasks seek for finding pairs of entities that are identical. However, real applications
might require searching for entity pairs satisfying a \emph{general binary relation} such as whether two items
are relevant to each other.

\begin{figure*}
    \centering
    \includegraphics[width=\textwidth]{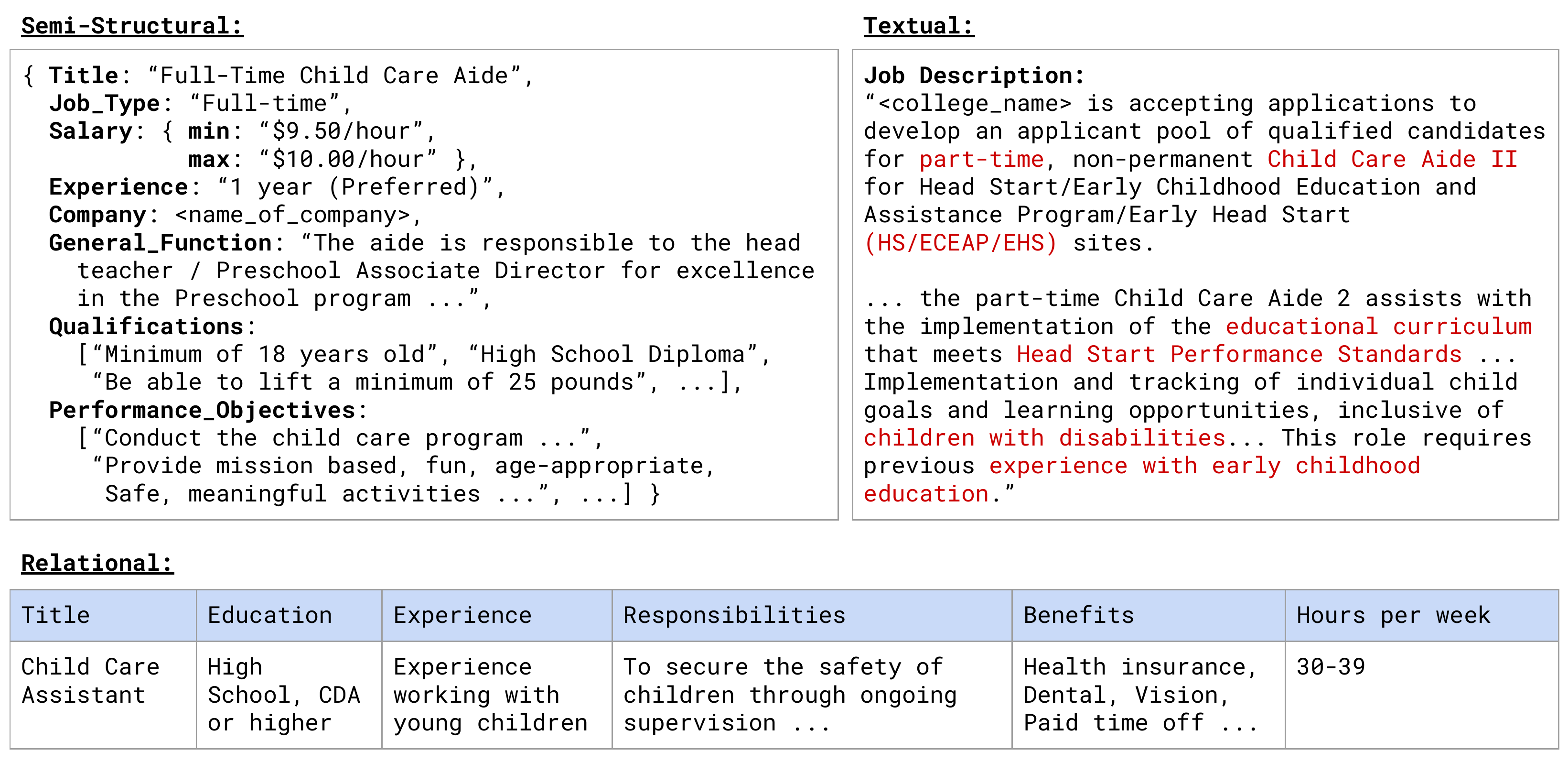}
    \caption{Real-world matching scenario from the job targeting applications. Job postings come in heterogeneous data format. Matching jobs of the same level and domain require both language and structural understanding of the entity records.}
    \label{fig:example}
\end{figure*}

\begin{example}
We illustrate the aforementioned new requirements of EM with a real scenario from 
job targeting platforms such as \url{indeed.com} in Figure~\ref{fig:example}. A common task that these platforms
need to perform is to \emph{find job postings representing jobs of the 
same level in the same domain}.
By doing so, they can support downstream applications such as
diversifying job search results or inviting candidates to apply (I2A).
The job postings naturally come in a variety of formats as companies can upload
plain text/PDF (upper-right) or use a template to create the job posting attribute-by-attribute (upper-left). 
Some datasets are in structural format (bottom) as they are collected by information extraction tools of the platform.
Because of the different data types, unifying their data schema is not possible.

The task is challenging as it requires the matching model to perform accurate
text matching such as comparing ``conduct child care program'' with ``experience working with young children'' or
``30-39 hours per week'' with ``full-time'' as well as understanding the document structure.
For example, the model needs to learn that the phrase 
``part-time'' in the textual document should be matched against 
the ``Job-Type'' JSON attribute.
\end{example}

In this work, as our first step, we formally define the problem
of \emph{Generalized Entity Matching} (GEM) given the above desiderata. 
GEM extends the classic EM problem setting by allowing matching entity records
of relational, semi-structural, or textual types. 
Besides, we also allow the matching relation to be a general binary relation customizable to a range of application needs. 
With the extended problem formulation, GEM covers an even wider range of applications
including matching JSON data, text matching~\cite{DBLP:conf/acl/DasS09,DBLP:conf/naacl/WilliamsNB18}, 
or data-driven fact checking~\cite{DBLP:journals/pvldb/Karagiannis0PT20}.

Next, to facilitate the development of matching models for addressing such generalized matching tasks,
we create \benchmark, a benchmark for evaluating models for GEM.
\benchmark consists of 7 real-world datasets covering various cases of interests
in matching data of different types.
We obtain the benchmark datasets by adapting and transforming existing EM datasets for structured data.
By doing so, \benchmark not only leverages the already existing ground truth labels, but also
covers application scenarios with different training set sizes, positive label ratios, and schema complexity etc.

We further conduct a profiling analysis by considering pairwise textual and structural similarity metrics
to help understand their difficulties for existing EM solutions that primarily rely on
global or attribute-wise similarity as features.
The profiling results indicate that existing EM solutions based on similarity measures cannot 
perform well on all \benchmark tasks.

To the best of our knowledge, \benchmark 
is the first benchmark for training and evaluating
entity matching models for heterogeneous data. 
We publish the \benchmark datasets at \url{https://github.com/megagonlabs/machamp}\footnote{Will be available soon}.

Our initial experiment on 7 popular learning-based EM solutions confirms the need for 
developing new matching techniques for GEM. Classic machine learning methods such as \rf and \svm
did not perform well in general.
Deep learning-based models based on pre-trained language models such as BERT~\cite{DBLP:conf/naacl/DevlinCLT19}
so far achieved the best results, but still they fail
to reach a moderate F1 score of 70\% in 3/7 of the \benchmark datasets.

\textbf{Contributions. } In summary, this paper makes the following contributions:
\begin{itemize}
\item We come up with the new research problem Generalized Entity Matching (GEM) based on the requirement of real world applications
which can capture matching data of different types such as structured, semi-structured,
or textual data. 
\item We release \benchmark, the first benchmark for GEM, 
for developing and evaluating novel matching solutions. 
By adapting and transforming existing EM datasets,
\benchmark takes advantage of the existing ground truth and covers a wide variety of real-world matching scenarios.
\item We conduct a profiling analysis with the help of pairwise similarity measures on the benchmark datasets.
The result of the analysis shows the difficulty of all tasks based on both textual and structural similarity.
\item We conduct an extensive set of experiments on 7 popular
EM methods including classical machine learning and deep learning approaches
based on pre-trained language models. 
Confirming the profiling analysis,
our initial results show that there is indeed significant room 
for improvement as existing methods fail to achieve a moderate level 
(i.e., 80\% average F1 score) of matching quality in some hard tasks.
\end{itemize}

The rest of the paper is organized as follows. We define the GEM problem and
task categorization in Section \ref{sec:tasks}. We present the \benchmark 
benchmark in Section \ref{sec:data}. 
Section \ref{sec:expset} and \ref{sec:expres} present the experiment
results and findings on existing EM solutions. 
Finally, we discuss related work in Section
\ref{sec:related} and conclude in Section \ref{sec:conclude}.

%% file: sec2-tasks.tex
\section{Tasks definitions} \label{sec:tasks}

In this section, we start by defining the GEM problem, i.e. the generalized version of the Entity Matching.
Next, we introduce 6 categories of practical scenarios that the \benchmark benchmark covers.
We also briefly introduce the benchmark datasets for classic EM from which \benchmark is constructed.

\subsection{Problem Formulation}\label{subsec-problem}

\newcommand{\attr}{\mathsf{attr}}
\newcommand{\val}{\mathsf{val}}

Given two tables of entities $E_A$ and $E_B$, the \emph{Entity Matching} (EM) problem aims at identifying all pairs of 
entity records $\langle e_a, e_b \rangle$ for $e_a \in E_A$, $e_b \in E_B$ that refer to the same real-world entity.
The tables $E_A$ and $E_B$ are called the left and right tables respectively for convenience.
In the classic setting~\cite{DBLP:journals/pvldb/KondaDCDABLPZNP16,DBLP:conf/sigmod/MudgalLRDPKDAR18,DBLP:journals/pvldb/0001LSDT20,DBLP:conf/edbt/BrunnerS20} of EM, it typically assumes that they are two relational tables having the same schema. 
Namely, there exists a relation $R(\attr_1, \dots, \attr_n)$ with $n$ attributes such that all records in 
$E_A$ and $E_B$ are elements of $R$.
We refer to this case \emph{structured} data matching with \emph{homogeneous} schema. 


To generalized the classic setting to more practical application scenarios,
we extend the definition as follows. We assume two base data types of \emph{number} and \emph{string}\footnote{We also include the list type but we omit it for sake of space.}. 
An entity record $e$ is a named tuple (i.e., a key-value collection) with a set of attributes 
$\{\attr_1, \dots, \attr_n\}$ where $\attr_i$ is the attribute name and $e.\attr_i$ denotes the attribute value.
Each value $e.\attr_i$ can be of either the base type or an entity record itself. In other words, we allow $e$ to be semi-structured data with
nesting attributes such as JSON. 
We define an entity table (or table for short) to be a set $E$ of entity records.
When an entity table $E$ contains an element with nested records, we call $E$ a semi-structured table.
Table $E$ is structured (or relational) otherwise.

When $n = 1$ and the only attribute $e.\attr_1$ is of the string type, we call the record $e$ an unstructured (or textual) record and $E$ an unstructured table.
Two entity tables are homogeneous if all the records share the same structure of attributes with respect to 
the attribute nesting. They are heterogeneous otherwise.
Following the above discussion, we can then formally define the GEM problem:

\begin{Problem}[GEM]
Given two structured, semi-structured, or unstructured entity tables $E_A$ and $E_B$
with homogeneous or heterogeneous schema, 
compute all record pairs $\langle e_a, e_b \rangle$ for $e_a \in E_A$, $e_b \in E_B$ where
$e_a$ and $e_b$ refer to the same real-world entity.
\end{Problem}

In our benchmark, we assume that within a structured and unstructured table, all records share the same schema; while in a semi-structured table, the schema of records can be different due to the inherited characteristics of semi-structured data. 

\smallskip
\noindent
\textbf{The matching task. } A standard entity matching pipeline
consists of two major steps: \emph{blocking} and \emph{matching}~\cite{DBLP:journals/pvldb/KondaDCDABLPZNP16}. 
The goal of the blocking step is to prune pairs of entity records
that are unlikely matches to avoid performing a full quadratic-size set of pairwise comparisons.
The matching step follows the blocking step to perform the actual pairwise comparison. 
In this paper, we focus our benchmarking effort on the matching step because
the recent development of matching techniques often relies on datasets for training
machine learning-based models~\cite{DBLP:journals/pvldb/EbraheemTJOT18,DBLP:conf/sigmod/MudgalLRDPKDAR18,DBLP:journals/pvldb/0001LSDT20}.
As such, it is tremendously important to get access to high-quality training and testing data for GEM.
Our goal is to provide the first benchmark for this purpose.
Formally, given two entity tables $(E_A, E_B)$,
a GEM dataset $D$ consists of a subset of record pairs from $E_A \times E_B$.
Every pair $\langle e_a, e_b \rangle \in D$ is associated with a ground truth label $y \in \{0, 1\}$
indicating whether $\langle e_a, e_b \rangle$ is a true match or not.
The dataset $D$ is typically split into the training, validation, and test sets for 
the model development and evaluation purpose.
Given a dataset $D$, the goal is to train a matching model $M$ such that for every pair $\langle e_a, e_b \rangle$,
$M(e_a, e_b) = 1$ if $e_a$ and $e_b$ are real match or 0 otherwise.

\subsection{Application scenarios}\label{subsec-task}

Based on the above discussion, compared with the classic EM problem, GEM covers a wider spectrum of matching scenarios and 
provides more flexibility.
By allowing different data formats with homogeneous or heterogeneous schema, 
GEM can support a variety of matching tasks with practical applications.
Next, we describe 6 such instantiations of GEM with the specific applications that they support.


\smallskip \noindent \textbf{Structured vs. Structured} \hspace{.5em} 
The most popular setting of classic EM is when both tables are structured and have homogeneous schema.
Classic EM has a wide range of applications from data integration~\cite{talburt2015entity} to knowledge base construction~\cite{DBLP:conf/acl/JiG11,DBLP:conf/cikm/WelchSD12}.
GEM further supports matching two structured tables with \emph{heterogeneous} schema.
By relaxing this constraint, we no longer require the schema alignment step before matching.

\smallskip \noindent \textbf{Semi-structured vs. Semi-structured}\hspace{.5em} In this scenario, both tables are semi-structured. 
This case is very popular in nowadays application as many data collections are organized in \emph{JSON} format. 
Due to the characteristics of semi-structured data, the schema of two semi-structured tables is not exactly the same in most cases.
Even within the same table, the schema of two records might also be different since if an attribute in a semi-structured record is missing, it will omit that attribute instead of having a NULL value as structured data did.
Therefore, this task is naturally heterogeneous in terms of data schema.
 
\smallskip \noindent \textbf{Semi-structured vs. Structured}\hspace{.5em} In this scenario, one of the tables is structured and another table is semi-structured.
In real-world applications, some data collections are stored in \emph{CSV} format while others are in \emph{JSON} format. 
To conduct entity matching between them, it requires not only to learn the semantic similarity between their textual content 
but also being aware of the structural information.

\smallskip \noindent \textbf{Structured vs. Unstructured}\hspace{.5em} In this scenario, one of the tables is structured and another table consists of free text.
The matching task will identify whether the description stated in the text is related to an entity in the structured table. 
A typical application of this task is claim verification~\cite{DBLP:journals/pvldb/Karagiannis0PT20} (a.k.a. fact check, fact verification) over text data based on the fact provided in the structured table. 

\smallskip \noindent \textbf{Semi-structured vs. Unstructured}\hspace{.5em} This setting is similar to the above one. 
The difference will be that the structured table is replaced with a semi-structured one. 

\smallskip \noindent \textbf{Unstructured vs. Unstructured}\hspace{.5em} This setting corresponds to 
the text matching task commonly appears in natural language processing (NLP) where both tables consist of free texts. 
There have already been many existing benchmark tasks in related research problems such as natural language inference~\cite{DBLP:conf/naacl/WilliamsNB18} and paraphrase identification~\cite{DBLP:conf/acl/DasS09}.
Researchers can directly utilize such resources in the NLP field to evaluate newly proposed techniques for this problem.
Therefore, here we will not include it in our benchmark.

\subsection{Data Collection and Processing}\label{subsec-preprocess}

In this paper, we build our benchmark tasks by leveraging the existing resources for EM over structured datasets.
By doing so, we can take advantage of the ground truth labels created 
in them and avoid expensive human labeling.
As they are originally proposed for matching relational data, we conduct necessary transformations to construct equivalent semi-structured or unstructured tables without changing the meaning of the records.
Specifically, the resources we rely on to create our benchmark tasks include the Magellan repository~\cite{DBLP:journals/pvldb/KondaDCDABLPZNP16}~\footnote{https://sites.google.com/site/anhaidgroup/useful-stuff/data}, \dm datasets~\cite{DBLP:conf/sigmod/MudgalLRDPKDAR18}~\footnote{http://pages.cs.wisc.edu/\~anhai/data1/deepmatcher\_data}, and the Web Data Commons collection (WDC)~\cite{DBLP:conf/www/PrimpeliPB19}~\footnote{http://webdatacommons.org/largescaleproductcorpus/v2/index.html}.
In the rest of the paper, we will call the two tables in a task left/right table when describing the dataset/task if there is no ambiguity.
Details of the resources are summarized as following:

First we introduce two resources from the Magellan repository.  
They consist of several smaller datasets.
The schema of tables in different datasets is also different.
Each dataset is stored in a CSV-like format, but many attributes are either missing or obviously unstructured (e.g. long review texts).
The labeled pairs in each dataset are not split into training/validation/test sets.
More details of them are introduced in Appendix A. 

\smallskip \noindent \textbf{Book}\hspace{.5em}  This series of tasks aims at matching books from different websites.
There are 5 datasets where the entities are books with the information such as title, author, publication, price, etc. 
The original data sources are from the websites Amazon, Barnes $\&$ Noble, and Goodreads. 

\smallskip \noindent \textbf{Movie}\hspace{.5em}  This series of tasks aims at identifying entity pairs that refer to the same movie.
There are 5 datasets where the entities are basic information of movies as well as the review comments.
The original data sources are from the websites Rotten Tomatoes, IMDB, TMD, Amazon, and Roger Ebert.

The following two resources are from the \dm datasets.
Different from the above ones, they have been well cleaned and formatted.
Besides, the pairs have been split into train, validation, and test sets ahead of time.

\smallskip \noindent \textbf{Restaurant}\hspace{.5em} 
The original dataset contains restaurant data from Fodors and Zagat.
It consists of two tables of the same schema 
$\{$name, addr, city, phone, type, class$\}$. 
The goal is to find the entities from different websites that refer to the same restaurant.
The number of pairs in the train/validation/test sets is 567, 190, and 189, respectively.

\smallskip \noindent \textbf{Citation}\hspace{.5em} This task is from the ER Benchmark~\cite{DBLP:journals/pvldb/KopckeTR10}, which is also used in \dm datasets.
The goal of the original task is to match bibliographic entities from different websites.
There are two datasets for this task: DBLP-ACM and DBLP-Google Scholar.
For DBLP-ACM, the number of pairs in the train/validation/test sets is 17223, 5742, and 5742, respectively.
For DBLP-Google Scholar, the number of pairs in the train/validation/test sets is 7417, 2473, and 2473, respectively.
For both datasets, the schema of both tables is $\{$title, authors, venue, year$\}$. 

Finally, we also look at the WDC datasets that are designed for large-scale product matching.

\smallskip \noindent \textbf{Products}\hspace{.5em} 
The original datasets consist of product records from multiple e-commerce websites.
The objective is to decide whether two entities refer to the same product.
There are four categories of tasks, namely Computer, Watch, Camera, and Shoe.
The tables in all datasets share the same schema where there are four attributes: \emph{title, description, brand, and specTable}.
Among them, the \emph{specTable} attribute consists of a series of key-value pairs and is somewhat semi-structured.
Meanwhile, the dataset is very sparse as the attributes Brand, and SpecTable are missing in most records.


%% file: sec3-datasets.tex
\section{The \benchmark benchmark} \label{sec:data}

To facilitate development matching solutions for GEM, we construct the \benchmark benchmark consisting of 7 datasets for training and testing
matching models. 
In this section, we first introduce the 7 datasets then present our analysis highlighting their hardness for existing EM solutions.

\subsection{Overview}\label{subsec-over}

We show a summary of \benchmark tasks in Table \ref{tab:stat1}.
Each task consists of two collections of entity records of possibly different formats
(i.e., relational, semi-structured, or textual). We denote the two collections by
the left and right tables respectively. 
When they are of the same type, they can have a homogeneous or heterogeneous schema. 
Table~\ref{tab:stat1} also summarizes the sizes and numbers of attributes
for each collection. For semi-structured data (row 2-4),
we report the number of attributes as the average number of leaf elements
in their JSON format. The numbers of attributes for structural or textual data
are constants.

We provide the training, validation, and test set for each dataset for 
model training and evaluation.
The size of the training set varies from a low 567 to over 17k covering both 
the low- and high-resourced settings in practice.
Similar to previous EM tasks, a GEM model should be trained on the training set,
tune its hyper-parameters on the validation set 
(e.g., selecting the best performing epoch), and report the precision, recall, and F1
scores on the test sets.

As discussed above, in this paper we will propose benchmark tasks for all but the last matching scenario (Unstructured vs. Unstructured).
Since there have been many benchmark tasks for Structured vs. Structured with homogeneous schema, we will also omit it and only prepare for the case with heterogeneous schema.
Besides, we also add an additional task for Semi-structured vs. Semi-structured where we force the two tables to be ``homogeneous''.
This is realized by filling the missing attributes with NULL values and ensure that all records have the same structure.

\setlength{\tabcolsep}{3pt}
\begin{table}[!ht]
\small
\caption{Dataset statistics.}\label{tab:stat1}
\begin{tabular}{ccccc|cccc} \toprule
           & \multicolumn{2}{c}{Left Table} & \multicolumn{2}{c|}{Right Table} & \multicolumn{4}{c}{Labeled Ground Truth} \\
           & \#row        & \#attr       & \#row        & \#attr       & Train    & Valid    & Test   & \% positive     \\ \midrule
\rrh   & 534          & 6.00         & 332          & 7.00         & 567      & 190      & 189    & 11.63\%   \\
\sso  & 2,616         & 8.65         & 64,263        & 7.34         & 17,223    & 5,742     & 5,742   & 18.63\%   \\
\ssh & 22,133        & 12.28        & 23,264        & 12.03        & 1,240     & 414      & 414    & 38.20\%   \\
\sr   & 29,180        & 8.00         & 32,823        & 13.81        & 1,309     & 437      & 437    & 41.64\%   \\
\stw & 9,234         & 10.00        & 9,234         & 1.00         & 5,540     & 1,848     & 1,846   & 11.80\%   \\
\stc & 20,897        & 10.00        & 20,897        & 1.00         & 12,538    & 4,180     & 4,179   & 14.07\%   \\
\rt  & 2,616         & 1.00         & 2,295         & 6.00         & 7,417     & 2,473     & 2,473   & 17.96\%  \\ \bottomrule
\end{tabular}
\end{table}

\subsection{Benchmark datasets}

\begin{figure*}
    \centering
    \includegraphics[width=1.0\textwidth]{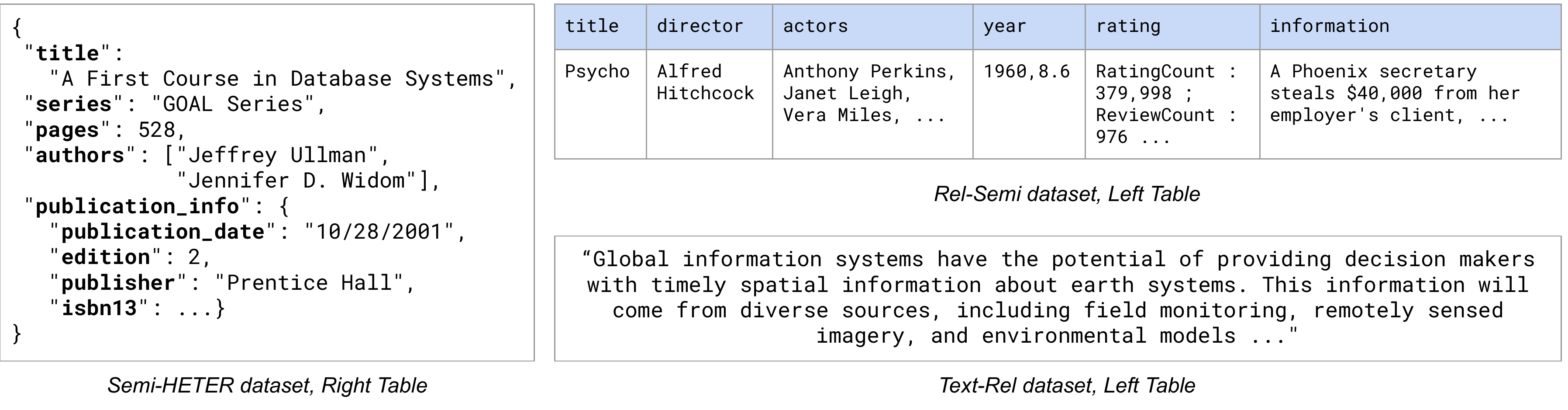}
    \caption{Examples of \benchmark dataset instances.}
    \label{fig:datasets}
\end{figure*}

The benchmark consists of the following 7 datasets. Examples of the datasets are shown in Figure \ref{fig:datasets}. When we name the datasets, 
we use ``\textsc{Rel}'' to denote whether the dataset contains a structured table,
``\textsc{Semi}'' to denote whether the dataset has a semi-structured (JSON)
collection, or ``\textsc{Text}'' if the dataset has textual data.
We omit the repetitive token when the two tables are of the same type.
We use ``\textsc{HOME}'' or ``\textsc{HETER}'' to denote whether the 
schema is homogeneous or heterogeneous. 
In the proposed datasets, tables with structured data are in CSV format; tables with semi-structured tables are in JSON format; while those with unstructured data are in text format.

\smallskip
\noindent
\textbf{\rrh. } This dataset is derived from the \textbf{Restaurant} task from \dm datasets. 
To convert the task into a case of heterogeneous schema, we modify the left table by combining ``addr'' and ``city'' into a single ``address'' attribute as well as 
combining ``type'' and ``class'' into a single ``category'' attribute.
To make the task more difficult to distinguish the power of different approaches, we drop the ``name'' attribute from the
right table. 
Finally, the schema of the left table is $\{$name, address, phone, category$\}$; while that of the right table is $\{$addr, city, phone, type, class$\}$. 

\smallskip
\noindent
\textbf{\sso. } We construct the second dataset from the DBLP-Scholar dataset from the \textbf{Citation} task.
As entities in the original dataset are in uniform relational tuples, we convert it into a task of matching semi-structured data with a homogeneous schema.
That is, we directly transform the structured records into semi-structured ones with key-value pairs.
The only difference is that the \emph{authors} attribute in the semi-structured record is a list generated by splitting the corresponding attribute into a list of names. 
We expect a matching solution to leverage the additional structural information to achieve improved matching quality.

\smallskip
\noindent
\textbf{\ssh. } We construct this task with semi-structured tables by combining 5 datasets in \textbf{Book} from the Magellan repository.
The 5 original datasets have different data sources thus naturally have different schemas. 
We group co-related attributes into nested tuples.
For example, the attributes \emph{publication$\_$date}, \emph{edition}, \emph{publisher}, \emph{isbn13} are grouped under a 
JSON key \emph{publication$\_$info}; while the attributes \emph{paperback$\_$price}, \emph{hardcover$\_$price}, \emph{nookbook$\_$price}, and  \emph{audiobook$\_$price} are grouped under a JSON key \emph{prices}.
For each record, we remove all attributes with \texttt{NULL} value to make the schema more flexible.
Thus as discussed before in Section~\ref{subsec-task}, the left and right tables in this task are naturally heterogeneous.

\smallskip
\noindent
\textbf{\sr. } We follow a similar approach to construct the \sr task by merging 5 datasets in \textbf{Movie} from the Magellan repository. 
For the left table, we make it as a structured table with unified schema $\{$ id, title, director, actors, year, rating, information $\}$.
In the original datasets, the attributes other than the first 6 ones mentioned here are merged as a string and regarded as \emph{information}.
For the right table, we treat it similarly as above in \ssh.
For example, we grouped related attributes such as $\{$rotten\_tomatoes, audience\_rating, reviews(list)$\}$ into nested attributes, e.g., ratings. 

\smallskip
\noindent
\textbf{\textsc{Semi-Text-w/c}.} The goal of these two datasets is to match
text with semi-structured data. We obtain 2 datasets from the WDC product
matching dataset. 
The records contain rich product information including both the textual product
title and product specifications in nested JSON format. 
To generate a meaningful matching task, we make the left table semi-structured and the right table unstructured in the following way:
For the left table, we simply keep only the \emph{specTable} attribute of the original record, which consists of a series of key-value pairs, as a semi-structured record.
For the right table, we concatenate the texts in \emph{title} and \emph{description} (if it is not empty) attributes of the original record as a piece of unstructured text.
To reach this goal, we only use the entity pairs where the \emph{specTable} attribute is not empty in the original dataset.
As the number of such pairs is limited in Camera and Shoe tasks in WDC, we only create two tasks \stw and \stc here, which are corresponding to the Watch and Computer sub-categories in the WDC dataset respectively.

\smallskip
\noindent
\textbf{\rt. } We construct a dataset for matching text and relational data  by extending the DBLP-ACM dataset from the \textbf{Citation} task.
The original dataset is a paper matching dataset similar to DBLP-Scholar above.
Instead of matching relational tuples, we replace the record in the left table (DBLP) with the abstract of the paper it refers to (or the first paragraph if the abstract is missing), which can be regarded as a high-level summary of the paper.
Then the task becomes a challenging matching task of matching the paper information (title, authors, venue, etc.) in the right table with paper abstracts in the left table.
Notice that here the meaning of matching is also changed: it is no longer ``refer to the same real-world entity'', but more generalized to ``entities with similar semantics''. 

For tasks \sso, \sr, and \textsc{Semi-Text-w/c}, the schema of different records might be different due to the inherited characteristics of semi-structured data.
The detailed schema of all tasks can be found in Appendix B. 

\subsection{Task Profiling}\label{subsec-profile}
 
Next, we define the profiling dimensions to analyze the proposed tasks.
To illustrate the characteristics that differentiate them from existing EM benchmark tasks, we focus on pairwise similarity analysis where we compute similarity scores between pairs of entity records or attributes.
We would like to understand whether existing approaches using popular similarity metrics can easily achieve high matching accuracy
on the new tasks. 
We follow the intuition that \emph{a feature is significant if it can easily \emph{separate} the positive and negative classes}.
Given a pairwise similarity function, we measure this separability by computing the gap between the \emph{average} pairwise similarity scores computed on the positive class and that on the negative class respectively. 
Intuitively, a large gap indicates that it is more likely for a similarity-based matching solution to achieve high accuracy for matching.
Since aligned schema is not available in most tasks of GEM, we cannot directly compute attribute-wise similarity. 
Instead, we consider two metrics by aggregating the similarity scores: \emph{textual} and \emph{structural} similarity.
In other words, we will regard them as the two profiling dimensions.

For the textual similarity, we employ Jaccard and Cosine similarity with TF-IDF, which are widely used matching features in both rule-based~\cite{DBLP:journals/tkde/Christen12}
and learning-based EM solutions~\cite{DBLP:journals/pvldb/KondaDCDABLPZNP16,DBLP:conf/sigmod/MudgalLRDPKDAR18,DBLP:journals/pvldb/EbraheemTJOT18}, as similarity functions. 
To this end, we first concatenate all attribute values into a single string for each entity.
Then for a pair of entities, we compute the similarity score as the similarity of the two flattened strings.

\SetKwInOut{Parameter}{Variables}
\begin{algorithm}[!ht]
	\KwIn{The first entity $e_a$ with attributes $A_1 \dots A_n$;
		the seccond entity $e_b$ with attributes $B_1 \dots B_m$;
		a string similarity function $\mathsf{sim}$ }
	\Parameter{similarity scores of elements -- $\mathsf{scores}$}
	\KwOut{aggregated similarity score}
	\If{$n = m = 1$}{
		\Return $\mathsf{sim}(e_a.A_1, e_b.B_1)$\;
	}
	\Else{
		\tcc{Identify shared attributes}
		$A_{\mathsf{shared}} \leftarrow \{A_1, \dots, A_n\} \cap \{B_1, \dots, B_m\}$\;
		$\mathsf{scores} \leftarrow \{\}$\;
		\For{all shared attributes $A \in A_{\mathsf{shared}}$}{
			\tcc{Recursion}
			$\mathsf{scores} \leftarrow \mathsf{scores} \cup  \{\text{StructuralSimilarity}(e_a.A, e_b.A, \mathsf{sim})\}$\;
		}
		\tcc{Flatten the remaining attributes}
		$\mathsf{str}_a \leftarrow \mathsf{flatten}(\{e_a.A | A \in \{A_1, \dots, A_n\} \setminus A_{\mathsf{shared}}\})$ \;
		$\mathsf{str}_b \leftarrow \mathsf{flatten}(\{e_b.A | A \in \{B_1, \dots, B_m\} \setminus A_{\mathsf{shared}}\})$ \;
		\tcc{Similarity of the two flattened strings}
		$\mathsf{scores} \leftarrow \mathsf{scores} \cup \mathsf{sim}(\mathsf{str}_a, \mathsf{str}_b)$\;
		\Return $\mathsf{avg(scores)}$\;
	}
	\caption{$\mathsf{Structural Similarity}$}
	\label{alg:structural}
\end{algorithm}	


For structural similarity, since the schema cannot be easily aligned in most tasks, we are not able to directly compute attribute-wise similarity. 
Thus we proposed an algorithm to compute and aggregate the similarity score recursively.
The detailed process is shown in Algorithm~\ref{alg:structural}.
When both entities are unstructured (with only one attribute), it directly returns the textual similarity (line: 2).
Otherwise, it will identify the shared attributes between two records (line: 4) and compute the structural similarity recursively:
For attributes unique to each record, we flatten those attributes (line: 8-9) and compute a single score of the two flatten strings (line: 10).
Finally, we aggregate all scores as the final results (line: 11) by averaging it over all pairs of attributes(including that between flatten strings computed in line 10).
Note that the structural similarity computed by Algorithm~\ref{alg:structural} de-generates to textual similarity when one side of the table is unstructured, or there is no common attribute between the two tables.

The analysis results are shown in Table~\ref{tab:stat2}.
As expected, the \rrh and \sso tasks have the largest average gaps between the positive and the negative class across the 4 similarity metrics.
That means they are relatively easy tasks.
This is reasonable because there is a relatively large overlap in their schema.
We conjuncture that existing EM methods are likely to achieve good results in these two tasks.
The \rt task has the smallest average similarity gaps. 
The reason might be that there is likely a consistent large content gap between an article abstract and a relational record.
Therefore,  its summary results in consistently low similarity scores.
The small similarity gap also indicates the difficulty of this task where existing EM solutions over structured tables might fail to achieve a moderate F1 score.

Note that the similarity gaps for textual similarity are larger than those for structural similarity (row 1-3).
This can be counter-intuitive at the first glance but is reasonable here because the pieces of useful matching information spread across multiple attributes. 
We conjuncture that in such cases, matching methods based on the entire record as context will perform
better than those based on local attribute-wise similarity.
As a result, it calls for new techniques that can capture structural similarity to better solve the GEM problem.

\setlength{\tabcolsep}{4pt}
\begin{table*}[!ht]
\caption{Statistics and similarity scores on the benchmark datasets.
Textual similarity is computed by flattening the both records.
Structural similarity is computed recursively over the shared structure
of the two records. We report the average similarity scores 
on both the positive and the negative class and their differences.
A larger gap indicates that the task is more likely to be solved
EM solutions relying on similarity scores as model features.}\label{tab:stat2}
\begin{tabular}{ccc|cc|ccc|ccc|ccc|ccc} \toprule
           & \multicolumn{2}{c|}{Left Table} & \multicolumn{2}{c|}{Right Table} & \multicolumn{3}{c|}{Textual Cosine} & \multicolumn{3}{c|}{Textual Jaccard} & \multicolumn{3}{c|}{Structual Cosine} & \multicolumn{3}{c}{Structural Jaccard} \\
           & \#char       & \#token      & \#char       & \#token      & neg        & pos       & diff      & neg        & pos        & diff      & neg        & pos        & diff       & neg         & pos         & diff       \\ \midrule
\rrh   & 79.21  & 14.35  & 14.35  & 11.33  & .0737 & .6995 & .6258 & .1226 & .5954 & .4727 & .0728 & .3035 & .2307 & .0828 & .2999 & .2171 \\
\sso  & 108.68 & 13.96  & 13.96  & 16.02  & .1703 & .6300 & .4597 & .1613 & .5411 & .3798 & .1464 & .5470 & .4006 & .1246 & .5116 & .3870 \\
\ssh & 305.04 & 43.65  & 43.65  & 23.85  & .2974 & .6065 & .3092 & .1750 & .3839 & .2090 & .2593 & .4976 & .2383 & .1535 & .3657 & .2122 \\
\sr   & 355.64 & 51.94  & 51.94  & 70.74  & .0241 & .0286 & .0045 & .0255 & .0305 & .0050 & .0241 & .0286 & .0045 & .0255 & .0305 & .0050 \\
\stw & 239.13 & 27.88  & 27.88  & 101.59 & .0962 & .1340 & .0378 & .0393 & .0406 & .0014 & .0962 & .1340 & .0378 & .0393 & .0406 & .0014 \\
\stc & 584.59 & 62.56  & 62.56  & 49.76  & .1204 & .2493 & .1289 & .0777 & .1143 & .0366 & .1204 & .2493 & .1289 & .0777 & .1143 & .0366 \\
\rt  & 866.44 & 127.70 & 127.70 & 21.00  & .0232 & .0241 & .0009 & .0307 & .0307 & .0000 & .0232 & .0241 & .0009 & .0307 & .0307 & .0000    \\ \bottomrule
\end{tabular}
\end{table*}

%% file: sec4-experiment.tex
\section{Experiment Setup} \label{sec:expset}

Next, we summarize the methods that we evaluated on \benchmark 
and the settings of our experiments.

\begin{table*}[ht]
	\caption{Main results (P: Precision; R: Recall; F: $F_1$ Score) of 7 popular learning-based EM methods.
	Classic machine learning methods such as \rf and \svm do not perform well in general. Deep learning methods 
	such as \transf, \sbert, and \ditto based on pre-trained language models achieve the best results so far.}\label{tbl:res}
	\scalebox{0.8}{
		\begin{tabular}{l|ccc|ccc|ccc|ccc|ccc|ccc|ccc}
			\toprule
			& \multicolumn{3}{c|}{\svm} & \multicolumn{3}{c|}{\rf} & \multicolumn{3}{c|}{\der} & \multicolumn{3}{c|}{\dm} & \multicolumn{3}{c|}{\transf} & \multicolumn{3}{c|}{\sbert} & \multicolumn{3}{c}{\ditto}  \\
			& P & R & F & P & R & F &P & R & F & P & R & F & P & R & F & P & R & F & P & R & F \\
			\midrule
			\rrh & 1.00 & 0.697 & 0.821 & 1.00 & 0.546 & 0.706 & 1.00 & 0.773 & 0.872 & 1.00 & 0.879 & 0.936 & 0.955 & 0.955 & 0.955 & 0.667 & 0.727 & 0.696 & 1.00 & 1.00 & \textbf{1.00} \\
			\sso & 0.604 & 0.473 & 0.53 & 0.917 & 0.630 & 0.747 & 0.894 & 0.858 & 0.875 & 0.890 & 0.832 & 0.861 & 0.938 & 0.939 & \textbf{0.938} & 0.856 & 0.893 & 0.874 & 0.947 & 0.916 & 0.931 \\
			\ssh & 0.839 & 0.164 & 0.274 & 1.00 & 0.151 & 0.262 & 0.617 & 0.182 & 0.282 & 0.358 & 0.245 & 0.291 & 0.907 & 0.308 & 0.460 & 1.00 & 0.535 & \textbf{0.697} & 0.846 & 0.484 & 0.616 \\
			\sr & 0.556 & 0.978 & 0.709 & 0.579 & 1.00 & 0.733 & 0.49 & 0.392 & 0.436 & 0.509 & 0.641 & 0.567 & 0.873 & 0.94 & 0.905 & 0.478 & 0.77 & 0.59 & 0.958 & 0.869 & \textbf{0.911} \\
			\stc & 0.579 & 0.537 & 0.557 & 0.902 & 0.508 & 0.650  & 0.756 & 0.289 & 0.418 & 0.766 & 0.311 & 0.442 & 0.89 & 0.883 & \textbf{0.886} & 0.851 & 0.751 & 0.798 & 0.822 & 0.813 & 0.818 \\
			\stw & 0.567 & 0.545 & 0.556 & 0.71 & 0.441 & 0.505 & 0.759 & 0.261 & 0.388 & 0.802 & 0.291 & 0.427 & 0.648 & 0.652 & \textbf{0.665} & 0.523 & 0.483 & 0.502 & 0.636 & 0.663 & 0.649  \\
			\rt & 0.49 & 0.392 & 0.436 & 0.603 & 0.26 & 0.363 & 0.727 & 0.416 & 0.529 & 0.784 & 0.404 & 0.534 & 0.616 & 0.646 & \textbf{0.631} & 0.372 & 0.295 & 0.329 & 0.656 & 0.601 & 0.627  \\
			\bottomrule
		\end{tabular}
	}
\end{table*}

\subsection{Methods Evaluated}

To the best of our knowledge, there are no previously proposed EM techniques that can be directly applied to the
GEM settings proposed in this paper.
Therefore, we extend some well-known entity matching methods for structured tables to evaluate the tasks here.
The results of such approaches could show the difficulty of proposed tasks and also set a lower bound of performance for more sophisticated matching methods proposed in the future.

\smallskip \noindent \textbf{Supported Vector Machine (SVM)} and \textbf{Random Forest} are two traditional machine learning techniques. 
They are widely used in text classification tasks before. 
They can be applied to the entity matching task by employing the bag of words representation for each training instance with bi-gram and tri-gram as the features.
 
\smallskip \noindent \textbf{\der}~\cite{DBLP:journals/pvldb/EbraheemTJOT18} employs the bi-directional Long Short Term Memory (LSTM) model to tackle the problem of entity matching. 
It first encodes each entity with an LSTM network and then fusion the representation of two entities to predict the match/unmatch result. 
  
\smallskip \noindent \textbf{\dm}~\cite{DBLP:conf/sigmod/MudgalLRDPKDAR18}  is an entity matching framework that uses Siamese RNN networks as the basic structure to aggregate the attribute values and then align the aggregated representations of the attributes. 
It proposes four matching models and the \textsf{Hybrid} one shows the best overall performance in the original paper.
   
\smallskip \noindent \textbf{\transf}~\cite{DBLP:conf/edbt/BrunnerS20} utilizes Transformer based models as the encoder for entity matching. 
It first transforms two entities into a sequence with special tags inserted before each attribute and concatenates the sequences of two entities as the input for the Transformer model.

\smallskip \noindent \textbf{\ditto}~\cite{DBLP:journals/pvldb/0001LSDT20} combines the pre-trained LMs with data augmentation techniques for entity matching. 
It can benefit from the inherited power of pre-trained LM and rely on the data augmentation techniques to address the problem of insufficient training instances.
Therefore, it achieves state-of-the-art performance in most benchmark tasks of entity matching.

\smallskip \noindent \textbf{\sbert}~\cite{DBLP:conf/emnlp/ReimersG19}  proposes a siamese architecture for pre-trained LMs for the task of sentence matching.
It first encodes two sentences with the same encoder separately and then concatenates the two representations and a vector generated from element-wise operation between them as the output for prediction.
It could also be applied to the task of entity matching.

\subsection{Serialization} \label{subsec-seq}

As existing approaches are designed for entity matching over structured data with homogeneous schema, we need proper serialization to evaluate them on the tasks proposed here.
The goal of serialization is to turn the original structured/semi-structured data into a token sequence that can be meaningfully ingested by each approach above while keeping as much structural information as possible.
To achieve this goal, we extend the serialization method proposed in \ditto~\cite{DBLP:journals/pvldb/0001LSDT20} and propose a reasonable way as follows.
How to develop new techniques for serialization and matching is out of the scope of this paper.

For structured tables, a data entry with $n$ attributes can be denoted as $e = \{ {\attr}_i, {\val}_i\}_{i \in [1,n]}$, where $\attr_i$ is the attribute name and $\val_i$ is the attribute value of the $i$-th attribute, respectively.
Then the serialization is denoted as 
\begin{center}
	{\em serialize(e)}: {\tt [COL]}\ $\attr_1$\ {\tt [VAL]}\ $\val_1$ $\ldots$ {\tt [COL]}\ $\attr_n$\ {\tt [VAL]}\ $\val_n$, 
\end{center}
where \textsf{[COL]} and \textsf{[VAL]} are two special tags indicating the start of attribute names and values respectively.
To serialize a pair of entities $\langle e, e'\rangle$, we concatenate them with a special token \textsf{[SEP]}.
For example, given the entity in the structured table shown in Figure~\ref{fig:datasets}, we serialize it as:
\begin{center}
\texttt{	[COL] title [VAL] Alfred Hitchcock  [COL] actors [VAL] Anthony Perkins, Janet Leigh... [COL] year [VAL] 1960,8.6 [COL] rating [VAL] RatingCount : 379,998 ;  ... [COL] information [VAL] A Phoenix secretary steals ...}
\end{center}

The semi-structured table can be serialized in a similar way. 
The only differences lie in: (i) for nested attributes, we recursively add the \textsf{[COL]} and \textsf{[VAL]}  tags along with attribute names and values in each level of nests; (ii) for attributes whose content is a list, we concatenate the elements in the list into one string separated by commas.
Given the semi-structured entity in Figure~\ref{fig:datasets}, we serialize it as:
\begin{center}
\texttt{	[COL] title [VAL] A First Course in Database Systems [COL] series [VAL] GOAL Series [COL] pages [VAL] 528 [COL] authors [VAL] Jeffrey Ullman, Jennifer D. Widom [COL] publication$\_$info [VAL] [COL] publication$\_$date [VAL] 10/28/2001 [COL] edition [VAL] 2 [COL] publisher [VAL] Prentice Hall [COL] isbn13 [VAL] .... }
\end{center}

For pre-trained LM-based approach, we further insert the token \textsf{[CLS]}, which is the special token necessary for BERT to encode the sequence pair into a feature vector which will be fed into the fully connected layers for classification, into the front of the sequence.
\begin{center}
	{\tt [CLS]} {\em serialize(e)} {\tt [SEP]} {\em serialize(e')} {\tt [SEP]}
\end{center}  

Moreover, we further customize the input sequence for different approaches as following:
\begin{compactitem}
	\item For \rf and \svm, we concatenate the two token sequences separated by a special tag \textsf{[ENTITY]} into one input sequence to the model.
	\item As \dm only accepts input with the same schema on two tables, we just make the schema of both tables as having only one attribute which is the token sequence generated above.
	\item Following the settings in~\cite{DBLP:conf/edbt/BrunnerS20}, we remove the special tag \textsf{[COL]} along with the attribute name and only keep the values as the input of \transf.
\end{compactitem}

\subsection{Environment and Settings}\label{subsec-setting}

All the experiments are conducted on a p3.8xlarge AWS EC2 machine with 4 V100 GPUs (1 GPU per run).
The pre-trained LMs are from the Transformers library~\cite{DBLP:conf/emnlp/WolfDSCDMCRLFDS20} and here we use BERT~\cite{DBLP:conf/naacl/DevlinCLT19} as the basic encoder for the three pre-trained LM-based approaches \transf, \ditto, and \sbert.
We use the base uncased variant of each model in all our experiments. 
To accelerate the training and prediction speed, we apply the half-precision floating-point (fp16) optimization.
We implement \rf and \svm using the Scikit-learn library and re-implement \der using the PyTorch platform. 
For other baseline methods \dm, \ditto, \transf and \sbert, we obtain the source code from the original open-sourced repositories.

We use Adam~\cite{DBLP:journals/corr/KingmaB14} as the optimizer for training and fix the batch size to be 16.
We tune the hyper-parameters by doing a grid search and select the one with the best performance.
Specifically, the learning rate is selected from \{$10^{-5}$, $3.0 \times 10^{-5}$, $5.0 \times 10^{-5}$\}.
The maximum sequence length for pre-trained LM-based methods is selected from \{128, 256, 384, 512\}; 
For the \dm approach, we use the default setting of its Hybrid model;
The number of training epochs is selected from \{5, 10, 20, 30, 40\}.
We use the $F_1$ score as the main evaluation metric and also report the values of precision and recall. 
For each run of experiments, we select the epoch with the highest $F_1$ on the validation set and report results on the test set.

\section{Results and Analysis}\label{sec:expres}

The results of evaluating existing approaches on the newly proposed benchmark tasks are shown in Table~\ref{tbl:res}.
The observations from such results are as follows.

Firstly, the results on different tasks are consistent with the difficulty of tasks profiled in Section~\ref{sec:data} in most cases.
Generally speaking, the results of \rrh and \sso have been already very promising even by extending existing solutions. 
For example, the results of \ditto on \rrh can reach a 1.0 in $F_1$ score, which is equivalent to that on the original task for homogeneous tables.
Meanwhile, those for tasks \ssh, \stw, and \rt are relatively low as the best approaches only achieved less than 0.7 in $F_1$ score. 
Such differences might be caused by the different complexity in the schema of data and 
textual similarity among positive and negative pairs.
More efforts are required to investigate the relationship between such factors and the performance to devise more effective approaches.

Secondly, the overall performance of pre-trained LM-based approaches is better than other ones.
Among all 7 tasks, \ditto achieved the best performance in 2 tasks; \sbert in 1 task, while \transf in the remaining 4 tasks.
The overall performances of \ditto and \transf are close with each other and much better than the other approaches.
The reason might be due to the superior ability of pre-trained LM in learning contextual word embedding, i.e. understanding the
semantics of tokens and discern homonyms and synonyms.
Moreover, pre-trained LMs are also robust to structural heterogeneity as they learn to pay attention to proper segments of two records when making matching decisions.
Therefore, it might be a promising starting point to utilize pre-trained LMs to address the problem of generalized entity matching.

Thirdly, the \rf and \svm approaches have the worst performance in most cases.
The main reason might be that as it just treats the problem in the same way as text classification and uses bag-of-words features, it fails to learn neither the structural information nor the matching between tokens or attributes.
Of course, the performance of \rf and \svm could be further improved via extensive efforts in feature engineering.
Nevertheless, that is out of the scope of this paper.

Finally, since some datasets are derived from original sources that are matching tasks between structured tables, we also make a comparison between the results here and the ones reported in the original paper.
Tasks \rrh and \sso are from \dm~\cite{DBLP:conf/sigmod/MudgalLRDPKDAR18}.
In task \rrh, the results of \dm and \ditto are both 1 in the original task; while here the result of \dm is 0.936. 
This is reasonable as less information is provided in our task than in the original one.
In task \sso, the results of \dm and \ditto are 0.947 and 0.956 respectively in the original task; while here the results are 0.861 and 0.931, respectively. 
The gap between them is not large because although the data becomes semi-structured in our task, the structure is not so complex and the two tables are homogeneous. 
For tasks \ssh and \sr, the datasets here are generated by combining several similar tasks from the Magellan repository.
The results on each original datasets are at least 0.9 using simple traditional machine learning approaches; while here the results drop significantly since both the structures of tables and the statistical information of the dataset significantly changes.
Therefore, they can be regarded as brand new tasks and it is meaningless to consider the results on original tasks.
Similarly, the available attributes or contents of \stc, \stw, and \rt are also very different from those in the corresponding original tasks.
Therefore, they should also be regarded as brand new tasks created by us.

From the above discussions, we can further make the following conclusions: (i) The coverage of \benchmark is sufficient to include different matching scenarios as well as various levels of difficulty; and (ii) there is still room for improving the performance on some difficult tasks like \ssh, \sr, and \rt by devising more advanced techniques to learn structural information rather than just inserting special tags in the process of serialization as we did here to extend existing solutions.

%% file: sec5-related.tex
\section{Related Work} \label{sec:related}

\subsection{Benchmark Tasks for Entity Matching}\label{subsec-benchmark}

Many previous studies have provided entity matching benchmark tasks that are publicly available.
The Cora Citation Matching dataset~\footnote{https://people.cs.umass.edu/\~mccallum/data/cora-refs.tar.gz} consisted of citations that might refer to the same paper.
The Leipzig DB Group Datasets~\cite{DBLP:journals/pvldb/KopckeTR10} provided data collections of bibliography and products that are from different websites~\footnote{https://dbs.uni-leipzig.de/research/projects/object\_matching/benchmark\_datasets\_
for\_entity\_resolution}.
Dude~\cite{draisbach2010dude} developed a toolkit for entity matching along with 3 small datasets.
The Magellan project~\cite{DBLP:journals/pvldb/KondaDCDABLPZNP16} provided several groups of entity matching tasks covering applications like restaurants, books, products, etc.
The Web Data Commons collection (WDC)~\cite{DBLP:conf/www/PrimpeliPB19} focused on large-scale entity matching and proposed four e-commerce related tasks.
Primpeli et al.~\cite{DBLP:conf/cikm/PrimpeliB20} systematically profiled existing benchmark tasks for entity matching and create rich features for them.
They focus on profiling existing tasks instead of creating new ones.
The recently proposed Alaska benchmark~\cite{DBLP:journals/corr/abs-2101-11259} aimed at evaluating data integration tasks in a unified manner, which also includes the entity matching task.
Meduri et al.~\cite{DBLP:conf/sigmod/Meduri0SS20} proposed the benchmark for entity matching with active learning, which targeted a different problem from our work.
All above benchmark tasks are designed for entity matching between relational datasets. 
Meanwhile, our newly proposed GEM benchmark aims at a broader scope of tasks where the tables on both sides can be structured, semi-structured, or unstructured.
We are aware that \dm~\cite{DBLP:conf/sigmod/MudgalLRDPKDAR18} proposes matching tasks with different data types, including structural, textual, and dirty.
The textual tasks in the context of \dm are the same as the unstructured vs. unstructured case mentioned in Section~\ref{sec:tasks}.
While the dirty tasks switch the contents of different attributes, they still require both tables to be relational and
have the same schema. We relax both restrictions in the problem definition of GEM.

\subsection{Entity Matching}\label{subsec-em}

There is a long line of studies about Entity Matching (EM) in the database and data mining community.
As surveyed in~\cite{DBLP:journals/csur/PapadakisSTP20}, there are two basic steps of EM: blocking and matching.
The blocking step aims at reducing the number of potential comparisons by clustering potentially matching entities into the same blocks 
to retain real matches as many as possible. The matching step performs pairwise comparisons 
within each block to identify matched entities. 

Many previous studies aimed at developing effective matching strategies, including rule-based and machine learning-based approaches~\cite{DBLP:journals/tkde/Christen12}.
Recently deep learning methods have been widely adopted in Entity Matching and achieve very promising results.
\der~\cite{DBLP:journals/pvldb/EbraheemTJOT18} first employed the Recurrent Neural Network (RNN) models to perform entity matching.
\dm~\cite{DBLP:conf/sigmod/MudgalLRDPKDAR18} combined the Siamese RNN model with alignment techniques to further improve the performance.
Kasai et al.~\cite{DBLP:conf/acl/KasaiQGLP19} developed the active learning method to deal with the situation where the number of training instances for entity matching is insufficient.
\textsf{MPM}~\cite{DBLP:conf/ijcai/FuHSCZWK19} proposed attribute comparison methods to improve the similarity measurement in the matching process.
\textsf{Seq2SeqMatcher}~\cite{DBLP:conf/cikm/NieHHSCZWK19} and \textsf{HierMatcher}~\cite{DBLP:conf/ijcai/FuHHS20} improved the performance of matching between heterogeneous data sources by applying additional alignment layers. 
However, they cannot be applied to more general scenarios such as when at least one table consists of semi-structured or unstructured data.
Some recent studies~\cite{DBLP:conf/edbt/BrunnerS20,DBLP:journals/pvldb/0001LSDT20,DBLP:conf/vldb/PeetersBG20} further adopted the pre-trained language models such as BERT for entity matching.
Brunner et al.~\cite{DBLP:conf/edbt/BrunnerS20} utilized the Transformer architecture as encoder and formulated the entity matching problem as sequence pair classification.
\sbert~\cite{DBLP:conf/emnlp/ReimersG19} proposed a Siamese Transformer framework for text matching tasks, which can also be utilized in the entity matching task. 
Peeters et al.~\cite{DBLP:conf/vldb/PeetersBG20} applied the BERT model in the application of products matching.
Our previous study \ditto~\cite{DBLP:journals/pvldb/0001LSDT20} integrated the pre-trained language models with data augmentation techniques and achieved the state-of-the-art performance.

%% file: sec6-conclude.tex
\section{Conclusion} \label{sec:conclude}

Entity Matching has been a popular task in many real-world applications.
In this paper, we generalize the research problem of entity matching based on real application scenarios by allowing the two tables of entities to be structured, semi-structured, or unstructured.
To facilitate evaluating this new problem, we create a benchmark \benchmark with 7 new tasks.
This is realized by transforming the tasks of existing benchmarks for EM between homogeneous structured records into those between tables 
with diverse structures.
We also reported detailed profiling results on these tasks and conducted an extensive set of experiments by adapting popular entity matching approaches for structured data.
From the evaluation results, we can see that the difficulty of the proposed tasks is consistent with those of our profiling efforts.
Besides, there is significant room for improvement for the performance of existing approaches by devising advanced techniques that jointly learn the text semantics and structural information.
We believe that our efforts in benchmarking will facilitate future research of entity matching for data of different structures by helping them develop and evaluate novel matching techniques. 

%% file: sec-schema.tex
\section{Schema of Original Datasets in Magellan}\label{app-orignal}

The schema of each dataset in \textbf{Book} and \textbf{Movie} is detailed as Table~\ref{tbl:book} and~\ref{tbl:movie}, respectively.
Here the column ``row'' means the row number of the list of datasets in the Magellan repository~\footnote{https://sites.google.com/site/anhaidgroup/useful-stuff/data$\#$TOC-The-784-Data-Sets-for-EM}.

\begin{table}[h]
	\small
	\caption{Statistics of Book Datasets}\label{tbl:book}
	\begin{tabular}{c|c|p{3cm}|p{3cm}} \toprule
		Row & \# Pairs & Left Table & Right Table \\ \hline
		15 & 374 & id, title, authors, pubyear, pubmonth, pubday, edition, publisher, isbn13, language, series, pages & Same as left \\ \hline
		16 & 396 & ID, Title, Description, ISBN, ISBN13, Page Count, First Author, Second Author, Third Author, Rating, Number of Ratings, Number of Reviews, Publisher, Publish Date, Format, Language, FileName & ID, Title, Author1, Author2, Author3, Publisher, ISBN13, PublicationDate, Pages, Product dimensions, Sales rank, Ratings count, Rating value, Paperback price, Hardcover price, Nookbook price, Audiobook price \\ \hline
		18 & 450 & ID, Title, Price, Author, ISBN13, Publisher, Publication\_Date, Pages, Dimensions & ID, Title, Used Price, New Price, Author, ISBN10, ISBN13, Publisher, Publication\_Date, Pages, Dimensions \\ \hline
		20 & 450 & ID, Title, Author, Price, Edition, ASIN, ISBN\_13, ISBN\_10, Paperback, Series, Publisher\_dummy, Publisher, Publication\_Date, Sold\_by, Language, Product Dimensions, Shipping Weight & ID, Title, Author, Hardcover, Paperback, NOOK\_Book, Audiobook, ISBN\_13\_DUMMY, ISBN\_13, Series, Publisher, Publication\_Date, Sales\_rank, Pages, Product\_dimensions \\ \hline
		22 & 398 & ID, Title, Author ,ISBN, Publisher, PublicationDate, Pages, price, ProductType  & ID, title, authors, cover, pages, publisher, language, ISBN-10, ISBN13, price \\ 
		\bottomrule
	\end{tabular}
\end{table}

\begin{table}[h!t]
	\small
	\caption{Statistics of Movie Datasets}\label{tbl:movie}
	\begin{tabular}{c|c|p{3cm}|p{3cm}} \toprule
		Row & \# Pairs & Left Table & Right Table \\ \hline
		3 & 600 &  Id, Name, Year, Release Date, Director, Creator, Actors, Cast, Language, Country, Duration, RatingValue, RatingCount, ReviewCount, Genre, Filming Locations, Description  &  Id, Name, YearRange, ReleaseDate, Director, Creator, Cast, Duration, RatingValue, ContentRating, Genre, Url, Description \\  \hline
		4 & 400 & ID, name, year, director, writers, actors & Same as left \\ \hline
		5 & 399 & ID, Title, Year, Rating, Director, Creators, Cast, Genre, Duration, ContentRating, Summary & Same as left \\ \hline
		6 & 412 & id, title, time, director, year, star, cost & id, title, time, director, year, star1, star2, star3, star4, star5, star6, rotten\_tomatoes, audience\_rating, review1, review2, review3, review4, review5 \\ \hline
		19 & 373 & id, movie\_name, year, directors, actors, critic\_rating, genre, pg\_rating, duration & id, movie\_name, year, directors, actors, movie\_rating, genre, duration \\ 
		\bottomrule
	\end{tabular}
\end{table}

\section{Detailed Schema of All Tasks}\label{app-all}

The schema of all tasks proposed in this paper are summarized as Table~\ref{tbl:schema}.
Note that for semi-structured data, the meaning of schema here is all the attributes that are possible to appear in a record.
If the value of an attribute is missing, the corresponding attribute will not appear in the record.

\begin{table}[h!t]
	\small
	\caption{The schema of all tasks}\label{tbl:schema}
	\begin{tabular}{c|c|p{2.4cm}} \toprule
		& Left Table & Right Table \\ \hline
		\rrh & name, address, phone, category & addr, city, phone, type, class\\ \hline
		\ssh & id, title, authors, venue, year & Same as left \\ \hline
		\sso & Varies  & Varies \\ \hline
		\sr & \begin{tabular}{c}
		id, title, director, actors, \\
		year, rating, information
		\end{tabular} & Varies \\ \hline
		\textsc{Semi-Text-w/c} & Varies & N/A \\ \hline
		\rt & N/A & id, title, authors, venue, year\\
		\bottomrule
	\end{tabular}
\end{table}

Among them, the terms ``Varies'' means that records in the corresponding table have different schema based on different sources they are from.
It happens in the tasks originated from the \textbf{Book} and \textbf{Movie} datasets in the Magellan repository as we create them by merging several different tasks.
Besides, as we use both Watch and Computer sub-categories from WDC, the \emph{specTable} of them are also different. 
Specifically, the schema of records in the left table of \ssh:
 
\begin{itemize}
	\item Book1: id, title, authors, pubyear, pubmonth, pubday, edition, publisher, isbn13, language, series, pages
	\item Book2: ID, Title, Description, ISBN, ISBN13, Page Count, First Author, Second Author, Third Author, Rating, Number of Ratings, Number of Reviews, Publisher, Publish Date, Format, Language, FileName
	\item Book3: ID, Title, Price, Author, ISBN13, Publisher, Publication\_Date, Pages, Dimensions
	\item Book4: ID, Title, Author, Price, Edition, ASIN, ISBN\_13, ISBN\_10, Paperback, Series, Publisher\_dummy, Publisher, Publication\_Date, Sold\_by, Language, Product Dimensions, Shipping Weight
	\item Book5: ID, Title, Author ,ISBN, Publisher, PublicationDate, Pages, price, ProductType
\end{itemize}	

The schema of records right tables in \ssh:
\begin{itemize}
	\item Book1: id, title, authors(list), publication\_info: [publication\_date, edition, publisher, isbn13], language, series, pages
	\item Book2: id, title, authors(list), publication\_info:[publication\_date, publisher, isbn13], pages, product\_dimensions, sales\_rank, ratings\_count, rating\_value, price:[paperback\_price, hardcover\_price, nookbook\_price, audiobook\_price]
	\item Book3: id, title, author(list), price:[used\_price, new\_price], publication\_info:[isbn10, isbn13, publisher, publication\_date], pages, dimensions
	\item Book4: id, title, author(list), price:[hardcover, paperback, nook\_book, audiobook], publication\_info:[isbn\_13\_dummy, isbn\_13, publisher, publication\_date], series, sales\_rank, pages, product\_dimensions
	\item Book5: id, title, authors(list), cover, pages, publisher, language, isbn:[isbn\_10. isbn13], price
\end{itemize}	

The schema of records in right tables in \sr:
\begin{itemize}
	\item Movie1: id, name, year\_range, release\_date, director, creator, cast, duration, rating:[rating\_value, content\_rating], genre, url, description
	\item Movie2: id, name, year, director, writers(list), actors(list)
	\item Movie3: id, title, year, director, creators, cast, genre, duration, rating:[rating, content\_rating], summary
	\item Movie4: id, title, time, director, year, stars(list), rating:[rotten\_tomatoes, audience\_rating, reviews(list)]
	\item Movie5: id, movie\_name, year, directors, actors, movie\_rating, genre, duration
\end{itemize}

The schema of records in left table in \textsc{Semi-Text}:
\begin{itemize}
	\item Watch: category, Merk, Product, Uitvoering, EAN, SKU, Tweakers ID
	\item Computer: Binding, Catalog Number List, Ean, Label, Manufacturer, Mpn, Part Number, Release Date, Studio, Upc, Color, Item Dimensions, Model, Package Dimensions, Product Group, Publisher, Size, Upc List
\end{itemize} 